\documentclass[12pt,a4paper]{article}
\usepackage{amsmath}
\usepackage[english]{babel}

\let\vphi\varphi

\let\p\partial

\let\ds\displaystyle

\begin{document}
%

\title{\Large\bf On application of Liouville type equations to constructing B\"acklund transformations}
\author{D K Demskoi \\[4mm] \small
School of Mathematics, University of New South Wales, \\
\small Sydney, NSW 2052, Australia \\
\small E-mail: demskoi@maths.unsw.edu.au
}

\label{firstpage}

\maketitle
\begin{abstract}
It is shown how pseudoconstants of the Liouville-type equations can be exploited as a tool for construction of the B\"acklund transformations.
Several new examples of such transformations are found. In particular we obtained the B\"acklund transformations for a pair of three-component analogs
of the dispersive water wave system, and auto-B\"acklund transformations for coupled three-component KdV-type systems.
\end{abstract}
\section{Introduction}
Since the discovery of integrability of the KdV equation several methods for classifying equations bearing the same property have been developed. The most well-known and fruitful of them are Painlev\'e test \cite{WB} and symmetry approach \cite{symmetry}. The common feature of these methods is that they provide only with necessary conditions of the integrability, though it appears that equations having passed these tests are integrable. This commonly can be proved by constructing the Lax representations or the B\"acklund transformations. One should mention that these two fundamental objects of soliton theory may be derived via truncation of Painlev\'e expansions, however, the calculations involved can be very tedious. 

The method presented in this paper can be successfully applied to constructing the B\"acklund transformation when the equation under consideration has a pair of hyperbolic Liouville-type equations \cite{LSS,ZhSok} as negative commuting flows.
%
%
The well known example is given by the hierarchy of sinh-Gordon--mKdV equations, used below to clarify the idea of the method. The method relies on the connection between the Miura and B\"acklund transformations on one hand, and the Miura transformations and the Liouville-type equations on the other. Thus the proposed method will complement existing approaches (see e.g. \cite{Rog1,Rog2,WTB,BP,RS,Matv}).

The paper is organized as follows. Below we rederive two well-known results  due to Wahlquist, Estabrook \cite{WB}, and Fordy \cite{fordy} concerned the B\"acklund transformations for the KdV, Sawada-Kotera, and Kupershimidt equations. In section~{\bf\ref{examples}} we construct the auto-B\"acklund transformations for several coupled  KdV-type systems recently presented in \cite{D3}. 

The auto-B\"acklund transformation for the KdV equation in the potential form
\begin{equation}
\vphi_\tau=\vphi_{xxx}+\frac{3}{2}\vphi_x^2 \\[3mm]
\label{KdV}
\end{equation}
is given by the relation \cite{WB}
\begin{equation}
\hat \vphi_x + \vphi_x=-\frac{1}{4}(\vphi-\hat\vphi)^2+2\lambda.
\label{BKdV}
\end{equation}
Here $\vphi, \hat \vphi$ are solutions of equation (\ref{KdV}), and $\lambda$ is an arbitrary parameter usually called "B\"acklund parameter". Equation (\ref{KdV}) is related to the potential mKdV equation
\begin{equation}
u_\tau=u_{xxx}-\frac{1}{2}u_x^3
\label{mKdV}
\end{equation}
by any of the Miura transformations
\begin{equation}
\vphi=\int\rho dx,\ \ \hat \vphi=-\int\hat\rho dx, \label{M1}
\end{equation}
where
\begin{equation}
\rho=u_{xx}-\frac{1}{2}u_x^2, \ \
 \hat\rho=u_{xx}+\frac{1}{2}u_x^2. \label{pscl2}
\end{equation}
Excluding variables $u_x,u_{xx}$ from (\ref{M1}) we obtain the auto-B\"acklund transformation for (\ref{KdV})
in the form
\begin{equation}
\hat \vphi_x + \vphi_x=-\frac{1}{4}(\vphi-\hat\vphi)^2.
\label{pBKdV}
\end{equation}
A B\"acklund parameter can be introduced into relation (\ref{pBKdV}) by applying the transformation 
$\vphi(x,t)\to x\lambda+\vphi(x+3\lambda t,t)$
corresponding to the classical symmetry $\varphi_\lambda=x+3t\varphi_x$ admitted by  equation (\ref{KdV}). 

The crucial point here is the way we get a couple of {\it different} Miura transformations relating equations (\ref{KdV}) and (\ref{mKdV}). 
First, we note that evolution equation (\ref{mKdV}) is the first higher symmetry (commuting flow) of the sinh-Gordon equation
\begin{equation}
u_{xt}=a e^u+b e^{-u}, \label{sh_G}
\end{equation}
where $a$, $b$ are arbitrary constants. Setting $b=0$ or $a=0$ we obtain the following couple of Liouville equations
\begin{equation}
u_{xt}=a e^u, \ \              
 u_{xt}=b e^{-u}
\label{Liouville12}
\end{equation}
for which functions (\ref{pscl2}) appear to be the simplest pseudoconstants, i.e. on solutions of corresponding equation (\ref{Liouville12}) they satisfy the characteristic equation 
$$\rho_t=0.$$
On the other hand a pseudoconstant of a Liouville-type equation determines a Miura transformation (see e.g. \cite{Sok, ZhSok}) for its higher symmetries. 

As it is evident from this example one need not make any assumptions on the order or the structure of the transformation to be found.
Applying this method we usually go in opposite direction: we start from an integrable hyperbolic equation, for example (\ref{sh_G}), find all its degenerate counterparts -- the Liouville type equations, and then find hierarchies of evolution equations related by theirs pseudoconstants. On the last step we construct (auto-)B\"acklund transformation excluding variables $u_i$. It is important for applicability of the method to have a couple of the Liouville-type equations generated by initial hyperbolic equation. 

Of course considered example is not a single case where the suggested procedure works. However, for some equations we get B\"acklund-type transformations without a parameter. One of such examples is related to  Tzitzeica equation
\begin{equation}
u_{xt}=a e^u+b e^{-2u}.
\label{Tz}
\end{equation}
Its first higher symmetry (commuting flow) has the form
\begin{equation}
u_\tau=u_{xxxxx}+5 (u_{xxx} u_{xx}- u_{xxx} u_x^2-u_x u_{xx}^2)+u_x^5.
\label{TzS}
\end{equation}
Setting $b=0$, and $a=0$ in (\ref{Tz}) we obtain the couple of Liouville equations
\begin{equation}
u_{xt}=a e^{u}, \ \ u_{xt}=b e^{-2u}
\label{LTz}
\end{equation}
having the following pseudoconstants correspondingly
\begin{equation}
\omega=u_{xx}-\frac{1}{2}u_x^2, \  \ \hat\omega=u_{xx}+u_x^2.
\label{psctz}
\end{equation}
On the solutions of equation (\ref{TzS}) functions
\begin{equation}
\vphi=\int \omega dx, \ \ \hat\vphi=\int \hat\omega dx
\label{MTz}
\end{equation}
satisfy the Kaup-Kupershmidt and Sawada-Kotera equations in potential form
\begin{equation}
\vphi_\tau= \vphi_{xxxxx}+10\,\vphi_{xxx}\vphi_x+\frac{15}{2} \vphi_{xx}^2+\frac{20}{3} \varphi_x^3,
\label{mSK}
\end{equation}
\begin{equation}
\hat \vphi_\tau= \hat \vphi_{xxxxx}+5 \hat \vphi_{xxx}\hat \vphi_x+\frac{5}{3} \hat \vphi_x^3.
\label{SK}
\end{equation}
Excluding variables $u_x, u_{xx}$ from (\ref{MTz}) we obtain the B\"acklund transformation for solutions of (\ref{mSK}) and (\ref{SK})
in the form \cite{fordy}
\begin{equation}
\vphi_x+\hat \vphi_x=-\frac{2}{3}(\vphi-\hat \vphi/2)^2.
\label{BckSK}
\end{equation}
We must finally state that a B\"acklund parameter cannot be introduced into transformation (\ref{BckSK}), and by this reason in the sequel we restrict ourselves to sinh-Gordon like equations.
\section{New examples}\label{examples} In this section we consider Lagrangian systems $\delta L / \delta u^i=0$ with
\begin{equation}
L=\sum_{i,j}g_{ij}(u)u_x^i u_t^j+f(u),
\label{genlagr}
\end{equation}
where $g_{ij}$ is the metric tensor of the configurational space with the coordinates $u^i$.
Systems with Lagrangian of the form (\ref{genlagr}) are usually called $\sigma$--models.
They play an important role in quantum field theory and in the theory of magnetics.
In paper \cite{D0} (see also \cite{D2,D1,D3}) a classification of Lagrangians (\ref{genlagr}) in the three-dimensional reducible Riemann space such that corresponding field systems admit higher polynomial symmetries is given. Below we consider the subclass of systems with
\begin{equation}
L=u_x u _t/2+\psi v_x w_t+a v^k e^{\lambda u}+b w^k e^{-\lambda u},
\label{Lagrs}
\end{equation}
where $\psi=1/(vw+c)$, and $c,\lambda,k,a,b=const$. It is assumed that $c\ne 0$, otherwise the configurational space is flat, and the problem reduces to the well-investigated case $\ds L=\sum\delta_{ij}u_x^iu_t^j+f(u)$, where $\delta_{ij}$ is the Kronecker's symbol. The choice of subclass (\ref{Lagrs}) is motivated by the fact that in this case the suggested procedure leads to a derivation of the auto-B\"acklund transformation with the B\"acklund parameter.
There are three integrable cases in (\ref{Lagrs}):
\begin{eqnarray}
& \lambda=\sqrt{2},\ \ k=1, \label{a} \\
& \lambda=k=1, \label{b} \\
& \lambda=k=2 \label{d}.
\end{eqnarray}
Note that the sinh-Gordon--mKdV hierarchy is a reduction of hierarchies considered here for $v=\mbox{const},\ \ w=\mbox{const}$. It is shown in \cite{D3} that hyperbolic systems corresponding to (\ref{a})-(\ref{d})  under condition $a=0$ or $b=0$ are of the Liouville-type.
Corresponding sets of pseudoconstants are given ibidem.
Using the method presented in the introduction it is easy to compute the B\"acklund transformations for all these cases. Below we give a detailed derivation of the B\"acklund transformation for case (\ref{a}). Calculations for cases (\ref{b}), (\ref{d}) are similar, but slightly more complicated, and we omit them presenting the final result only.
\\ {\bf Case (\ref{a}).} Hyperbolic system corresponding to (\ref{a}) and its simplest higher symmetry are
\begin{equation}
\begin{array}{l}
u_{xt}=\sqrt{2}(ave^{\sqrt{2}u}-bwe^{-\sqrt{2}u}), \\[1mm]
v_{xt}=be^{-\sqrt{2}u}\psi^{-1}+\psi v_xv_t w,\ \ w_{xt}=ae^{\sqrt{2}u}\psi^{-1}+w_xw_t v\psi,
\end{array}
\label{sysa}
\end{equation}
\begin{equation}\begin{array}{l}
u_{\tau}=\sqrt{2}\psi v_x w_x,\ \ v_{\tau}=v_{xx}-2v\psi v_x w_x +\sqrt{2} u_x v_x,\\[1mm] 
w_{\tau}=- w_{xx} + 2w \psi v_x w_x +\sqrt{2} u_x w_x.
\label{sym2}
\end{array}\end{equation} 
If $b=0$, then the complete set of pseudoconstants of (\ref{sysa}) is given by
\begin{eqnarray}
\rho = (\sqrt{2}\,u_{xx}-u_x^2-2v_x w_x \psi)/6, \label{psc11} \\
\textstyle \theta = v_{xx} v_x^{-1}+\frac{\sqrt{2}}{2}\, u_x-\psi w_x v,\label{psc12} \\
\vphi = \psi v_x (w_{xx}-v_x w_x \psi w-\sqrt{2} u_x w_x)\label{psc13}.
\end{eqnarray}
It follows from the form of Lagrangian (\ref{Lagrs}) that pseudoconstants corresponding to the case $a=0$ can be obtained from (\ref{psc11})-(\ref{psc13}) by the substitution $u_i\to -u_i, v_i\to w_i, w_i\to v_i$. They are of the form 
\begin{eqnarray}
\hat \rho=-(\sqrt{2} u_{xx}+u_x^2+2 v_x w_x \psi)/6,\label{psc21}\\
\textstyle\hat \theta =w_{xx}w_x^{-1}-\frac{\sqrt{2}}{2}\, u_x-\psi v_x w, \label{psc22}\\
\hat \vphi = \psi w_x (v_{xx}-v_x w_x \psi v+\sqrt{2} u_x v_x).\label{psc23}
\end{eqnarray}
Relations (\ref{psc11})-(\ref{psc13}) and (\ref{psc21})-(\ref{psc23}) determine differential substitutions of system (\ref{sysa}) into the following systems correspondingly
\begin{equation}
\begin{array}{l}
m_\tau=\frac{2}{3}\vphi,\\[2mm]
n_\tau=n_{xx}+n_x^2+3 m_x, \\[2mm]
\vphi_\tau=-\vphi_{xx}+2(n_x\vphi)_x,
\end{array} \qquad
\begin{array}{l}
\hat m_\tau=-\frac{2}{3}\hat \vphi,\\[2mm]
\hat n_\tau=-\hat n_{xx}-\hat n_x^2-3 \hat m_x, \\[2mm]
\hat \vphi_\tau=\hat \vphi_{xx}-2(\hat n_x \hat \vphi)_x,
\end{array}
\label{pmsys}
\end{equation}
where $m_x=\rho, n_x= \theta$. It is easy to see that systems (\ref{pmsys}) are related by the discrete transformation $\tau\to-\tau$. The integrability of (\ref{pmsys}) was established in \cite{D3} by constructing its bi-Hamiltonian structure.
There it was also pointed out that (\ref{pmsys}) can be obtained from the Yajima-Oikawa system \cite{YO} in the same way as the dispersive water wave system (the Kaup-Broer system) from the NLS equation. Thus systems (\ref{pmsys}) can be regarded as three-component analogs
of the dispersive water wave system.

To obtain the B\"acklund transformation for (\ref{pmsys}) we exclude variables $u_i, v_i,w_i$ from relations (\ref{psc11})-(\ref{psc13}) and (\ref{psc21})-(\ref{psc23}). From relations (\ref{psc11}), (\ref{psc21}) we obtain
\begin{equation}
u_x=\frac{3}{\sqrt{2}}\int(\rho-\hat \rho)\,dx, \ \
\label{f1}
\end{equation}
\begin{equation*}
\psi\,v_x w_x=-\frac{3}{2}(\hat\rho+\rho)-\frac{9}{4}\Big(\int(\hat\rho-\rho)dx\Big)^2.
\label{f2}
\end{equation*}
On the other hand, it follows from (\ref{psc12}), (\ref{psc22}) that
\begin{equation}
\psi\,v_x w_x=\exp\Big(\int(\hat \theta+\theta)dx\Big).
\label{f3}
\end{equation}
Thus the first relation of the sought transformation is
\begin{equation}
\hat\rho+\rho=-\frac{2}{3}\exp\Big(\int(\hat \theta+\theta)dx\Big)-\frac{3}{2}\Big(\int(\hat\rho-\rho)dx\Big)^2.
\label{prebckl1}
\end{equation}
By expressing $v_{xx}$ from relation (\ref{psc12}) and substituting it into (\ref{psc23}) we get
\begin{equation}
\hat \vphi=\frac{\sqrt{2}}{2}\,\psi v_x w_x(\sqrt{2}\theta+u_x).
\label{f4}
\end{equation}
Substituting expressions (\ref{f1}) and (\ref{f3}) into (\ref{f4}), we obtain
\begin{equation}
\hat \vphi=\Big(\theta+\frac{3}{2}\int(\rho-\hat \rho)dx\Big)\exp\Big(\int(\theta+\hat\theta)dx\Big).
\label{prebckl2}
\end{equation}
Similarly from relations (\ref{psc13}), (\ref{psc22}), (\ref{f1}) and (\ref{f3}) we have
\begin{equation}
\vphi=\Big(\hat \theta-\frac{3}{2}\int(\rho-\hat \rho)dx\Big)\exp\Big(\int(\theta+\hat\theta)dx\Big).
\label{prebckl3}
\end{equation}
Thus relations (\ref{prebckl1}), (\ref{prebckl2}) and (\ref{prebckl3}) represent the B\"acklund transformation for solutions of systems (\ref{pmsys}). Passing on to potentials $(\rho,\theta) \to (m_x,n_x), (\hat \rho,\hat \theta) \to (\hat m_x,\hat n_x)$ 
we bring the B\"acklund transformation to the form
\begin{equation}
\begin{array}{l}
m_x =-\hat m_x-\frac{3}{2}(m-\hat m)^2-\frac{2}{3}\,e^{\hat n+n}+\lambda, \\[2mm]
n_x = \frac{3}{2} (\hat m-m)+\hat \vphi\,e^{-n-\hat n}, \ \
\vphi = e^{\hat n+n} (\hat n_x+\frac{3}{2} (\hat m- m)).
\end{array}
\label{bckla}
\end{equation}
Here the parameter $\lambda$ is injected into (\ref{bckla}) by applying the transformations
\begin{equation*}
\begin{array}{l}
m \to -x\lambda+m(x+\lambda,t), \ \ n \to -3t\lambda+n(x+\lambda,t), \ \ \vphi\to\vphi(x+\lambda,t), \\[2mm]
\hat m \to x\lambda+\hat m(x+\lambda,t), \ \ \hat n \to -3t\lambda+\hat n(x+\lambda,t), \ \ \hat\vphi\to\hat\vphi(x+\lambda,t),
\end{array}
\label{trans1}
\end{equation*}
generated by the classical symmetries of (\ref{pmsys})
\begin{equation*}
\begin{array}{l}
m_\lambda=m_x-x,\ \ n_\lambda= n_x-3t,\ \ \vphi_\lambda=\vphi_x, \\[2mm]
\hat m_\lambda=\hat m_x+x,\ \hat n_\lambda=\hat n_x-3t,\ \ \hat\vphi_\lambda=\hat\vphi_x .
\end{array}
\end{equation*}
{\bf Case (\ref{b}).} The hyperbolic system corresponding to the case (\ref{b}) takes the form
\begin{equation}
\begin{array}{l}\label{sysb}
u_{tx}=2a v^2e^{2u}-2b\,w^2\,e^{-2u},\\[2mm] 
v_{tx}=2bw\,\psi^{-1} e^{-2u}+\psi\,w\,v_t\,v_x,\ \
w_{tx}=2av\,\psi^{-1} e^{2u}+\psi\,v\,w_t\,w_x.
\end{array}
\end{equation}
In the degenerate cases $a=0$ or $b=0$ system (\ref{sysb}) possesses complete sets of pseudoconstants. In particular if $b=0$ then the pseudoconstants are
\begin{equation}
\begin{array}{l}
\rho = u_2-u_1^2-2 v_1 w_1 \psi,\ \ \theta = v_2v_1^{-1}+u_1-v \psi w_1,\\[2mm]
\varphi = -3 \psi (2 v_1^3 w_1 w^2 \psi^2-2 \psi w w_2 v_1^2 +\psi  v w_1 w_2 v_1-2 \psi vv_1 w_1^2  u_1\\[2mm]
\quad-2 \psi v_1^2 w_1^2-2 v_1 w_1 u_2+2 v_2 u_1 w_1-v_2 w_2-4 w_2 u_1 v_1+4 v_1 w_1 u_1^2\\[2mm]
\quad+4 w\psi v_1^2 w_1  u_1+v_1 w_3).
\end{array}\label{PsCB1}
\end{equation}
In the case $a=0$ we have
\begin{equation}
\begin{array}{l}
\hat \rho = -u_2-u_1^2-2 v_1 w_1 \psi,\ \ \hat\theta = w_{xx}w_x^{-1}-u_x-w \psi v_1,\\[2mm]
\hat\varphi = -3 \psi (2 w_1^3 v_1 v^2 \psi^2-2 \psi v v_2 w_1^2 +\psi  w v_1 v_2 w_1+2 \psi ww_1 v_1^2  u_1\\[2mm]
\quad-2 \psi v_1^2 w_1^2+2 v_1 w_1 u_2+2 w_2 u_1 v_1-v_2 w_2+4 v_2 u_1 w_1+4 v_1 w_1 u_1^2\\[2mm]
\quad-4 v\psi w_1^2 v_1  u_1+w_1 v_3).
\end{array}\label{PsCB2}
\end{equation}
The latter is obtained from (\ref{PsCB1}) by the substitution $u_i\to-u_i, v_i\to w_i,w_i\to v_i$.
The simplest higher symmetry of (\ref{sysb}) is of the third order and have the following explicit form
\begin{equation}
\begin{array}{l}
u_\tau=\frac{1}{4}u_{xxx}-\frac{3}{2}(w_{xx}v_x-v_{xx}w_x)\psi+\frac{3}{2}(wv_x-vw_x)v_xw_x\psi^2\\
\qquad +3\psi u_xv_xw_x-\frac{1}{2}u^3_x,\\[2mm]
v_\tau=v_{xxx}+3v_{xx}(u_x-vw_x\psi)+\frac{3}{2}u_{xx}v_x+3v^2v_xw^2_x\psi^2-3v^2_xw_x\psi \\[1mm]
\qquad  +\frac{3}{2}v_xu^2_x-6vu_xv_xw_x\psi,\\[2mm]
w_\tau=w_{xxx}-3w_{xx}(u_x+wv_x\psi)-\frac{3}{2}u_{xx}w_x+3w^2w_xv^2_x\psi^2 -3w^2_xv_x\psi \\[1mm]
\qquad   +\frac{3}{2}w_xu^2_x+6wu_xv_xw_x\psi.                                               \label{symb}
\end{array}\end{equation}
Sets of pseudoconstants (\ref{PsCB1}) and (\ref{PsCB2}) determine the Miura transformations of system (\ref{symb}) into the system
\begin{equation}
\begin{array}{l}
m_\tau=\frac{1}{4} m_{xxx}+\frac{3}{4} m_x^2+\varphi, \\[1mm]
n_\tau = n_{xxx}+\frac{3}{2} (2 n_x n_{xx}+n_x m_x)+n_x^3+\frac{3}{4}m_{xx}, \\[1mm]
\varphi_\tau=\varphi_{xxx}+(\varphi n_{xx}-n_x \varphi_x+\varphi n_x^2+\varphi m_x)_x-\frac{3}{2}\varphi_x m_x,
\end{array}\label{trans3B1}
\end{equation}
where $m_x=\rho, n_x=\theta$. 
%
The auto-B\"acklund transformation for system (\ref{trans3B1}) is therefore
\begin{equation}
\begin{array}{l}
m_x = -\hat m_x-\frac{1}{2}(m-\hat m)^2-4 e^{n+\hat n}+\lambda , \\[2mm]
n_{xx} = 3 e^{n+\hat n}-\frac{1}{3}\hat \varphi\, e^{-n-\hat n}+\frac{1}{4}(m-\hat m)^2 \\[2mm]
\qquad +\frac{1}{2}(\hat m-m-2 n_x)(n_x-\hat n_x)+\hat m_x-\lambda/2, \\[2mm]
\varphi = -\frac{3}{4}(m-\hat m)^2 e^{n+\hat n}-3 e^{2n+2\hat n} \\[2mm]
\qquad -\frac{3}{2} e^{n+\hat n}(2\hat n_{xx}+2\hat m_x+(n_x-\hat n_x)(m-\hat m-2\hat n_x)-\lambda).
\end{array}\label{bcklb}
\end{equation}
{\bf Case (\ref{d}).} The hyperbolic system takes the form
\begin{equation}
\begin{array}{l}\label{sysd}
u_{tx}=a\,v\,e^u-b\,w\,e^{-u},\\[2mm]
v_{tx}=b\,\psi^{-1} e^{-u}+\psi\,w\,v_t\,v_x,\ \ w_{tx}=a\,\psi^{-1} e^u+\psi\,v\,w_t\,w_x.
\end{array}
\end{equation}
Setting $b=0$ in (\ref{sysd}) we get the Liouville-type system with the pseudoconstants
\begin{equation}
\begin{array}{l}
\rho = 2 u_{xx}-u_x^2-2 v_x w_x \psi,\ \ \theta = v_{xx} v_x^{-1}-w_x v \psi+u_x,\\[2mm]
\varphi =  v_x \psi (-w_{xxx}-2w^2 \psi^2 w_xv_x^2-v w \psi^2v_x w_x^2 +2\psi w v_x  w_{xx}-2 w_x u_x^2 \\[2mm]
\quad-3\psi w v_x w_x  u_x+\frac{3}{2}\psi w_x^2  v_x+w \psi v_{xx} w_x +w_x u_{xx}+3 u_x w_{xx}).
\end{array}\label{PsCD1}
\end{equation}
The other possibility $a=0$ gives us a system with the following pseudoconstants
\begin{equation}
\begin{array}{l}
\hat \rho = -2 u_{xx}-u_x^2-2 v_x w_x \psi,\ \ \hat \theta = w_{xx} w_x^{-1}-v_x w \psi-u_x,\\[2mm]
\hat \varphi =  w_x \psi (-v_{xxx}-2v^2 \psi^2 v_xw_x^2-v w \psi^2w_x v_x^2 +2\psi v w_x  v_{xx}- v_x u_{xx} \\[2mm]
\qquad+3\psi v w_x v_x  u_x+\frac{3}{2}\psi v_x^2  w_x+v \psi w_{xx} v_x -3 u_x v_{xx}-2 v_x u_x^2).
\end{array}\label{PsCD2}
\end{equation}
The simplest higher symmetry of system (\ref{sysd}) is
\begin{equation}
\begin{array}{l}
u_\tau =- \frac{1}{2}u_{xxx}+\frac{3}{2}\,\psi(v_{xx}w_x-v_x w_{xx})+\frac{1}{4}u^3_x+\frac{9}{2}\,\psi u_x v_x w_x\\[2mm]
\qquad+\frac{3}{2}\,\psi^2 v_x w_x(v_x w- v w_x),\\[2mm]
v_\tau =v_{xxx}+\frac{3}{2}\,u_{xx} v_x +3 v_{xx}(u_x-\psi v w_x)+\frac{9}{4}\,u_x^2 v_x- 6\psi v u_x v_x w_x\\[2mm]
\qquad +3\,\psi v_x w_x( \psi v^2 w_x- \frac{1}{2}v_x),\\[2mm]
w_\tau = w_{xxx}-\frac{3}{2}\,u_{xx} w_x -3 w_{xx}(u_x+\psi v_x w)+\frac{9}{4}\,u_x^2 w_x\\[2mm]
\qquad + 6\psi w u_x  v_x w_x+3\,\psi v_x w_x(\psi w^2 v_x- \frac{1}{2}w_x).                                 \label{symd}
\end{array}
\end{equation}
Miura transformations (\ref{PsCD1}) and (\ref{PsCD2}) relate system (\ref{symd}) with the system 
\begin{equation}
\begin{array}{l}
m_\tau=-\frac{1}{2}m_{xxx}+6 \varphi-\frac{3}{8} m_x^2, \ \ n_\tau = n_{xxx}+3 n_x n_{xx}+\frac{3}{4} m_x n_x+n_x^3,\\[2mm]
\varphi_\tau=\varphi_{xxx}+3\,(n_x^2 \varphi-n_x \varphi_x)_x+\frac{3}{4}\, m_x \varphi_x
\end{array}\label{trans3D1}
\end{equation}
where $m_x=\rho, n_x=\theta$. Excluding variables $u_i,v_i,w_i$ from relations (\ref{PsCD1}) and (\ref{PsCD2}) we obtain the auto-B\"acklund transformation for system (\ref{trans3D1})
\begin{equation}
\begin{array}{l}
m_x = -\hat m_x-\frac{1}{8}(m-\hat m)^2-4e^{n+\hat n}-4\lambda,\\[2mm]
n_{xx} = \frac{1}{4}n_x (\hat m- m)-n_x^2-\hat \varphi\,e^{-n-\hat n}+\frac{1}{2}e^{n+\hat n}+\lambda, \\[2mm]
\varphi = \frac{1}{2}e^{2n+2\hat n}+e^{n+\hat n}\left(\frac{1}{4}\hat n_x (m- \hat m)- \hat n_x^2-\hat n_{xx}+\lambda\right),
\end{array}\label{bckld}
\end{equation}
where $\lambda$ is the B\"acklund parameter.

One of applications of the B\"aklund transformations is generation of exact solutions from a given ones. Consider for example the simplest case: system (\ref{pmsys}) and its B\"acklund transformation (\ref{bckla}). Take $m=n=\vphi=0$ as a seed solution, then relations (\ref{bckla}) yield
the following system of ordinary differential equations
\begin{equation*}
\hat m = -\frac{2}{3}\hat n_x,\ \ \hat \vphi =\p_x e^{\hat n}, \ \ \hat n_{xx}-\hat n_x^2-e^{\hat n}+\frac{3}{2}\lambda=0.
\end{equation*}
Solving this system we get expressions for functions $\hat m,\hat n,\hat\vphi$ containing two arbitrary functions of time, the latter can be found after substituting these expressions into the second system (\ref{pmsys}). This gives one soliton solution of (\ref{pmsys}) which can be represented as
\begin{equation}
\begin{array}{l}
\ds \hat m=\frac{2}{3}\lambda \frac{\kappa_1e^{\lambda (t\lambda+2 x)}+\kappa_2 e^{\lambda^2 t}}{\kappa_1e^{\lambda (t\lambda+2 x)}-\kappa_2e^{\lambda^2 t}+2 e^{\lambda x} }, \\[4mm]
\ds \hat n=\lambda x-\log\left(2 e^{\lambda x}+\kappa_1 e^{\lambda (t\lambda+2 x)}-\kappa_2 e^{\lambda^2 t}\right)+\log(2 \lambda^2), \\[3mm]
\ds \hat \vphi=2 \lambda^3 \frac{\left(\kappa_1 e^{\lambda (t \lambda+2 x)} +\kappa_2 e^{\lambda^2 t}\right) e^{\lambda x}}{\left(\kappa_1 e^{\lambda (t \lambda+2 x)}-\kappa_2 e^{\lambda^2 t}+2 e^{\lambda x}\right)^2},
\end{array}
\label{}
\end{equation}
where $\kappa_1,\kappa_2=\mbox{const}$.
%
\section{Acknowledgments and concluding remarks}
In this paper we have presented the method which allowed us to construct B\"acklund transformations for several three-component evolution systems presented in \cite{D3}. One of these systems is closely related to the Yajima-Oikawa system, the B\"acklund transformation for which was discussed in \cite{CSZ} from the viewpoint of Lax pair. We would like to point out that the Lax pairs for (\ref{trans3B1}) and (\ref{trans3D1}) are of the fourth order and have quite complicated structure \cite{D1}, so it would be difficult to find corresponding B\"acklund transformations if we tried to do that using methods previously known.
%

The author is grateful to Prof. V.V. Sokolov for useful discussions, and to Prof. Q.P.~Liu for pointing out the reference \cite{CSZ}.
\\
\label{lastpage}
\end{document}